\documentclass[conference]{IEEEtran}
\ifCLASSINFOpdf
\else
\fi
\usepackage{graphicx} % For graphics
\usepackage{fixltx2e} % For subscript in regular text
\usepackage{subfig}
\usepackage[dvipsnames]{xcolor}
\usepackage{tikz}
\usepackage{standalone}
\usepackage{array}
\usepackage{multirow}
\usepackage{bigstrut}
\usepackage[para]{threeparttable}
\usepackage{units}
\usepackage{gensymb}

\begin{document}

\title{An Application-Specific VLIW Processor with Vector Instruction Set for CNN Acceleration}

\author{\IEEEauthorblockN{Andreas Bytyn, Rainer Leupers and Gerd Ascheid}
\IEEEauthorblockA{Institute for Communication Technologies and Embedded Systems, RWTH Aachen University\\
Email: bytyn@ice.rwth-aachen.de}}

%\author{}


%Telephone: (800) 555--1212, Fax: (888) 555--1212}


\newcommand\copyrighttext{%
  \footnotesize \textcopyright 2019 IEEE. Personal use of this material is permitted. Permission from IEEE must be obtained for all other uses, in any current or future media, including reprinting/republishing this material for advertising or promotional purposes, creating new collective works, for resale or redistribution to servers or lists, or reuse of any copyrighted component of this work in other works.}
\newcommand\copyrightnotice{%
\begin{tikzpicture}[remember picture,overlay]
\node[anchor=south,yshift=10pt] at (current page.south) {\fbox{\parbox{\dimexpr\textwidth-\fboxsep-\fboxrule\relax}{\copyrighttext}}};
\end{tikzpicture}%
}

\maketitle

\copyrightnotice

\begin{abstract}
In recent years, neural networks have surpassed classical algorithms in areas such as object recognition, e.g. in the well-known ImageNet challenge. As a result, great effort is being put into developing fast and efficient accelerators, especially for Convolutional Neural Networks (CNNs). In this work we present ConvAix, a fully C-programmable processor, which – contrary to many existing architectures – does not rely on a hard-wired array of multiply-and-accumulate (MAC) units. Instead it maps computations onto independent vector lanes making use of a carefully designed vector instruction set.

%The presented core is capable of executing up to 192 MAC operations per cycle at a target clock frequency of 400 MHz in 28nm CMOS, thereby offering state-of-the-art performance with proper flexibility within its target domain. Simulation results for several 2D convolutional layers from well known CNNs (AlexNet, VGG-16) show an average core utilization of XX \% while achieving state-of-the-art energy efficiency of up to YYY GOP/s/W using 16-bit fixed-point arithmetic. A comparison to other recently published designs shows that our core can achieve competitive energy efficiency while offering higher area efficiency and throughput.

The presented processor is targeted towards latency-sensitive applications and is capable of executing up to 192 MAC operations per cycle. ConvAix operates at a target clock frequency of 400 MHz in 28nm CMOS, thereby offering state-of-the-art performance with proper flexibility within its target domain. Simulation results for several 2D convolutional layers from well known CNNs (AlexNet, VGG-16) show an average ALU utilization of 72.5\% using vector instructions with \mbox{16 bit} fixed-point arithmetic. Compared to other well-known designs which are less flexible, ConvAix offers competitive energy efficiency of up to 497 GOP/s/W while even surpassing them in terms of area efficiency and processing speed.
\end{abstract}


% \fi
\IEEEpeerreviewmaketitle

\section{Introduction}
Since their introduction to the broad public, Convolutional Neural Networks (CNNs) have been adopted for many tasks such as object classification and detection \cite{Krizhevsky2012} \cite{Redmon2015}. Their ability to extract meaningful features out of data has been the key enabling factor for their superior performance compared to other approaches. However, this comes at the price of increased computational complexity, specifically the number of Multiply-And-Accumulate (MAC) operations can reach well into the several GMAC per frame as summarized in \cite{Sze2017}. Thankfully, there is a lot of explicit parallelism contained in CNNs, thereby offering many options for accelerating them with domain-specific hardware architectures.

The most time- and energy-consuming computational kernel of every CNN is the 3D-convolution of so called input feature maps (e.g. RGB input for the first layer) with multiple sets of filters that compute layer-wise output feature maps \cite{Sze2017}. The focus of this paper is therefore on the acceleration of these convolutions. As described in \cite{Ma2017} and \cite{Chen2017}, the number of off-chip memory accesses and the management of the on-chip memories both play a detrimental role for energy consumption and arithmetic utilization within the accelerator. There are many parameters for the convolutions that affect the efficiency and it is the designers task to find a suitable trade-off between them.

Important parameters to consider are for example the order in which convolutions are executed as well as the tiling of input and output feature maps into e.g. column- and depth-slices. The optimal choice of these parameters however is dependent on the specific CNN model, thereby making it desirable to have some degree of flexibility in the data flow. The authors believe that an Application-Specific Instruction Set Processor (ASIP), as presented in this paper, can offer a decent trade-off between flexibility and efficiency. Some parameters, such as unrolling-factors that result in hardware parallelism, must be decided at design time. However, other parameters such as tiling-factors and loop-order, can be flexibly adjusted in software. Since a fully featured C/C++-compiler is generated for our ASIP automatically, computational kernels can easily be re-used or adapted by means of software libraries.

In Section \ref{sec_related_work} a brief overview of existing hardware accelerators for CNNs is given. Afterwards, the convolutional kernel is introduced in Section \ref{sec_cnn} and an overview of the processor architecture is given in Section \ref{sec_architecture}. Section \ref{sec_results} presents post place-and-route results of the ASIP implemented in a 28nm CMOS technology, as well as some relevant benchmarks running the state-of-the-art CNNs AlexNet and VGG-16. Finally, Section \ref{sec_conclusion} concludes the paper.

\section{Related Work}
\label{sec_related_work}
Many of the published accelerators are comprised of a large array of processing entities (PEs) with some application-specific interconnectivity between them. These accelerators often offer immense performance in terms of throughput (GOP/s) and energy efficiency (GOP/s/W), yet lag the desired flexibility when it comes to the employable data flow patterns and the on-chip data management.
In \cite{Chen2017a}, the authors present a 12x14 MAC array that aims to maximize data reuse (and therefore minimize off-chip accesses) by applying a specific computation scheme called \textit{row stationary}. Some data flow flexibility is achieved by subdividing the 2D MAC array into slices and distributing parallel computations amongst these slices. This flexibility has its limits though, as only a pre-defined set of parameters can be adapted at runtime.
In \cite{Moons2017a}, a C-programmable processor is presented that makes use of a 16x16 MAC array to accelerate convolutions. The MAC array is supplied with data by a RISC processor, which gives flexibility in terms of the on-chip data management, however the MAC array itself does not offer any additional flexibility.
Both \cite{Chen2017a} and \cite{Moons2017a} apply voltage scaling for their circuit to demonstrate the potential energy efficiency improvement when operating at a lower voltage and clock frequency.

Furthermore, since many CNNs can be quantized down to \mbox{8 bit} fixed-point \cite{Lin2015} \cite{Moons2016a}, existing architectures exploit this by either designing their MAC units to be narrow to begin with (e.g. \mbox{12 bit} as in \cite{Cavigelli2015a}), or by applying techniques such as precision-gating or using subword parallelism at runtime \cite{Moons2017a}.
Other architectures such as \cite{Azarkhish2018} and \cite{Gao2017} attempt to alleviate the memory bottleneck in CNNs by employing near-memory computation and using 3D memory.

\section{CNN Data Flow}
\label{sec_cnn}
In general, CNNs consist of a number of concatenated layers, each executing a pre-defined operation on a 3- or 4-dimensional tensor (depending on whether batch processing is applied), thereby generating an output tensor that is used as input to the following layer. The most common layers include the convolutional layer, max-pooling layer for tensor-downsampling, and recently also so-called depth-wise and point-wise convolutional layers \cite{Howard2017}, which are special cases of the regular convolutional layer. For the remainder of this paper we will focus on the convolutional layers, as they constitute most of the computational expenditure in modern CNNs. Furthermore, batch processing is not considered since the processor presented here is intended for real-time applications that are latency-sensitive.

\begin{figure}[h]
\vspace{-0.1cm}
\centering
\includegraphics[scale=1.0]{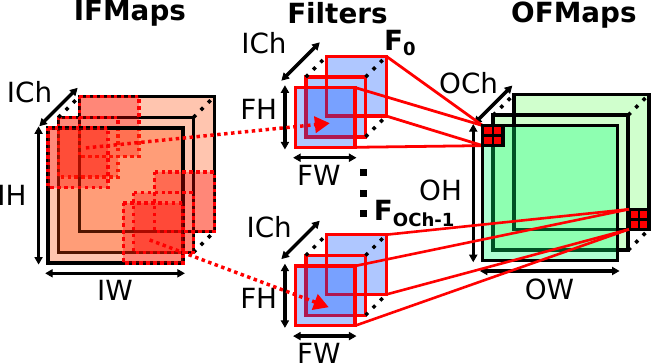}
\caption{Convolutional layer example.}
\label{fig_convlayer}
\vspace{-0.1cm}
\end{figure}

Fig. \ref{fig_convlayer} illustrates the 3D-convolution operation used in CNNs. A volume of input feature maps (IFMaps) consisting of \textit{ICh} separate channels (also called the depth), each of them being of dimension \mbox{\textit{IH} x \textit{IW}}, is convolved with \textit{OCh} banks of filters (\mbox{\textit{F\textsubscript{0}}..\textit{F\textsubscript{OCh-1}}}). In this process, each filterbank creates one output volume of dimension \mbox{\textit{OH} x \textit{OW}} by convolving each IFMap with the corresponding filter of dimension \mbox{\textit{FH} x \textit{FW}} and accumulating the results of the different filters.

As mentioned before, not all IFMaps, OFMaps and filters can be kept in on-chip memory at the same time, so only subsets of the data are available for processing. This can be interpreted as slicing the input and/or output tensors into smaller chunks of data for which parallel processing is possible. For more information on the different slicing options, the interested reader is referred to \cite{Ma2017} and \cite{Chen2017}.

\begin{figure}[h]
\vspace{-0.1cm}
\centering
\includegraphics[scale=1.0]{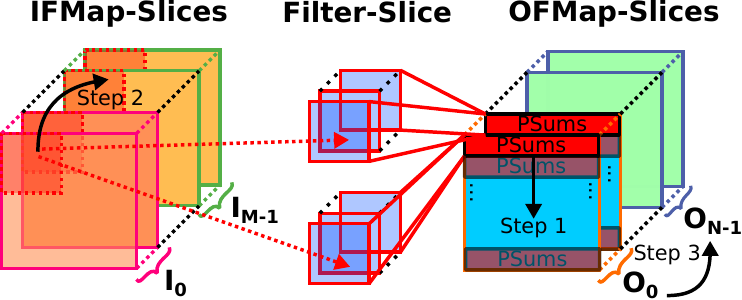}
\caption{Exemplary data flow used for benchmarks in Sec. \ref{sec_results}.}
\label{fig_conv_dataflow}
\vspace{-0.1cm}
\end{figure}

Due to its software programmability, ConvAix supports a variety of slicing options. Fig. \ref{fig_conv_dataflow} illustrates one option that is particularly suitable for networks such as AlexNet and \mbox{VGG-16}, which is also used for the benchmarks presented in Section \ref{sec_results}.

Both IFMaps and OFMaps are sliced along their depth dimension to build M input slices \textit{I\textsubscript{0}}..\textit{I\textsubscript{M-1}} as well as N output slices \textit{O\textsubscript{0}}..\textit{O\textsubscript{N-1}}. Each output slice is then processed in a row-wise fashion (step 1) in order to re-use existing IFMap rows when shifting the filter window to the next row. For each slice, filters are pre-loaded before processing starts, while IFMap-rows and OFMap-rows are loaded and stored concurrently on demand. Partial sums (PSums) of the incomplete OFMaps are accumulated in local scratchpad memories and only if necessary buffered in off-chip memory, which also happens concurrent to processing. After all IFMaps of the current slice \mbox{\textit{I\textsubscript{m}}} have been processed, the next slice \mbox{\textit{I\textsubscript{m+1}}} is loaded (step 2). Finally, the current OFMap slice \mbox{\textit{O\textsubscript{n}}} is complete and the next slice \mbox{\textit{O\textsubscript{n+1}}} can be processed (step 3). Note that if the IFMaps are not sliced along their depth-dimension, no intermediate off-chip buffering of PSums is required.

\section{Processor Architecture}
\label{sec_architecture}

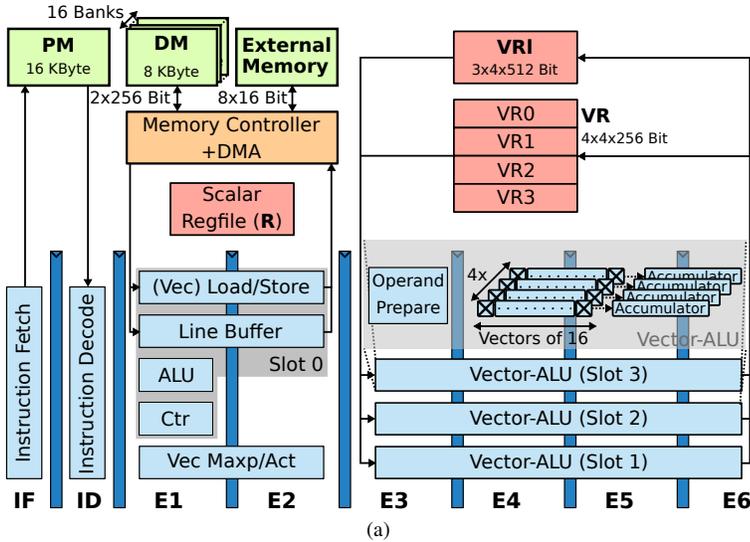
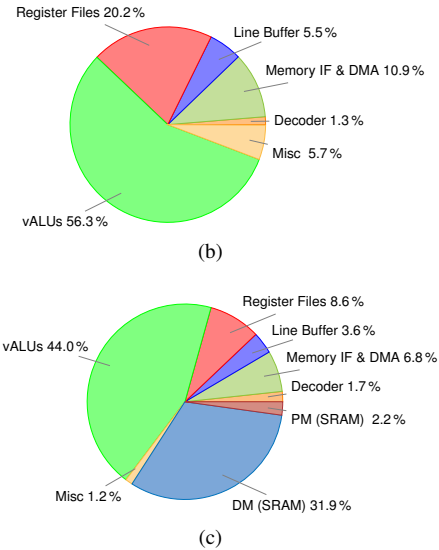
\begin{figure*}[!t]
\vspace{-0.35cm}
\begin{tabular}{cc}
    \multirow{-7}[2]{*}{\subfloat[]{
        \includegraphics[scale=0.9]{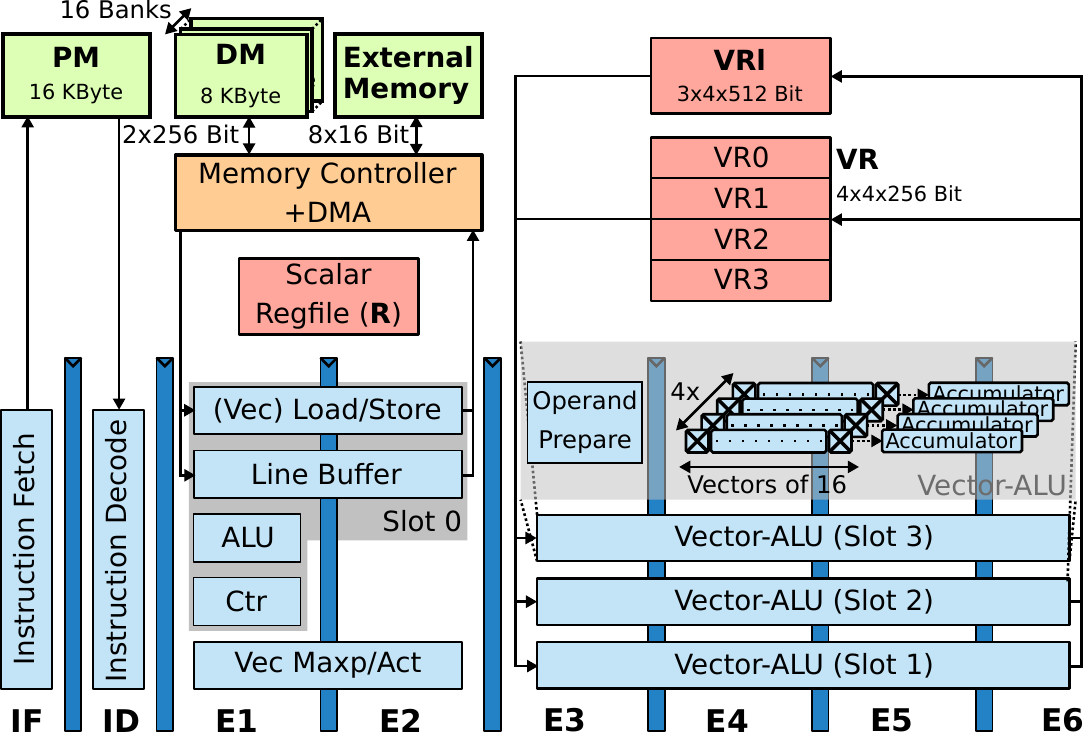}
        \label{fig_core_architecture} %
    }} & 
    \subfloat[]{
        \def\angle{0}
\def\radius{2.5}
\def\lineradius{2.0}
\def\cyclelist{{"orange","LimeGreen","blue","red","green", "Dandelion", "RoyalBlue", "Maroon"}}
\newcount\cyclecount \cyclecount=-1
\newcount\ind \ind=-1
\begin{tikzpicture}[nodes = {font=\sffamily}]
  \foreach \percent/\name in {
      1.3/Decoder,
      10.9/Memory IF \& DMA,
      5.5/Line Buffer,
      20.2/Register Files,
      56.3/vALUs,
      5.7/Misc
    } {
      \ifx\percent\empty\else               % If \percent is empty, do nothing
        \global\advance\cyclecount by 1     % Advance cyclecount
        \global\advance\ind by 1            % Advance list index
        \ifnum7<\cyclecount                 % If cyclecount is larger than list
          \global\cyclecount=0              %   reset cyclecount and
          \global\ind=0                     %   reset list index
        \fi
        \pgfmathparse{\cyclelist[\the\ind]} % Get color from cycle list
        \edef\color{\pgfmathresult}         %   and store as \color
        % Draw angle and set labels
        \draw[fill={\color!50},draw={\color}] (0,0) -- (\angle:\radius)
          arc (\angle:\angle+\percent*3.6:\radius) -- cycle;
%         \node at (\angle+0.5*\percent*3.6:0.7*\radius) {\percent\,\%};
        \node[pin=\angle+0.5*\percent*3.6:\name$\;$\percent\,\%]
          at (\angle+0.5*\percent*3.6:\lineradius) {};
        \pgfmathparse{\angle+\percent*3.6}  % Advance angle
        \xdef\angle{\pgfmathresult}         %   and store in \angle
      \fi
    };
\end{tikzpicture}
        \label{fig_core_area}
    } \\
 &  \subfloat[]{
         \def\angle{0}
\def\radius{2.5}
\def\lineradius{2.0}
\def\cyclelist{{"orange","LimeGreen","blue","red","green", "Dandelion", "RoyalBlue", "Maroon"}}
\newcount\cyclecount \cyclecount=-1
\newcount\ind \ind=-1
\begin{tikzpicture}[nodes = {font=\sffamily}]
  \foreach \percent/\name in {
      1.7/Decoder,
      6.8/Memory IF \& DMA,
      3.6/Line Buffer,
      8.6/Register Files,
      44.0/vALUs,
      1.2/Misc,
      31.9/DM (SRAM),
      2.2/PM (SRAM)
    } {
      \ifx\percent\empty\else               % If \percent is empty, do nothing
        \global\advance\cyclecount by 1     % Advance cyclecount
        \global\advance\ind by 1            % Advance list index
        \ifnum7<\cyclecount                 % If cyclecount is larger than list
          \global\cyclecount=0              %   reset cyclecount and
          \global\ind=0                     %   reset list index
        \fi
        \pgfmathparse{\cyclelist[\the\ind]} % Get color from cycle list
        \edef\color{\pgfmathresult}         %   and store as \color
        % Draw angle and set labels
        \draw[fill={\color!50},draw={\color}] (0,0) -- (\angle:\radius)
          arc (\angle:\angle+\percent*3.6:\radius) -- cycle;
%         \node at (\angle+0.5*\percent*3.6:0.7*\radius) {\percent\,\%};
        \node[pin=\angle+0.5*\percent*3.6:\name$\;$\percent\,\%]
          at (\angle+0.5*\percent*3.6:\lineradius) {};
        \pgfmathparse{\angle+\percent*3.6}  % Advance angle
        \xdef\angle{\pgfmathresult}         %   and store in \angle
      \fi
    };
\end{tikzpicture}
         \label{fig_core_power}
     }
\end{tabular}
\caption{ConvAix instruction pipeline and storage overview (\ref{fig_core_architecture}), processor area breakdown (w/o SRAMs) (\ref{fig_core_area}) and exemplary power distribution (\ref{fig_core_power}) for AlexNet layer 3 (\mbox{8 bit} gated precision).}
\vspace{-0.35cm}
\end{figure*}

The architecture overview of the proposed ASIP, called ConvAix, is shown in Fig. \ref{fig_core_architecture}. It consists of 8 pipeline stages (ID, IF, E1..E6) with 4 heterogeneous VLIW issue-slots to exploit instruction-level parallelism. Slot 0 is reserved for control instructions as well as memory operations both for the on-chip as well as the off-chip memory. ConvAix also offers a scalar ALU that is used for address calculations and house-keeping computations, e.g. loop-counter updates. In addition to the regular load/store-unit, an application-specific line buffer is used to cache IFMap rows. Slots 1-3 each offer a pipelined SIMD vector datapath, whereas each datapath (vALU) itself again consists of 4 separate SIMD vector-slices that can be programmed in C using specific vector primitives added to the compiler. The vector parallelism is set to 16, resulting in a total of \mbox{4 x 16 = 64} MAC operations per issue-slot and \mbox{3 x 64 = 192} MAC operations in total for all 3 slots. Furthermore, slot 1 includes an application-specific unit that operates on single vectors of size 16, which is used for calculating activation functions and performing max-pooling.

In general, the ASIP uses \mbox{16 bit} fixed-point arithmetic for both the scalar ALU as well as the vector units. However the ALU in slot 0 also offers a \mbox{32 bit} datapath to perform computations for addressing larger memories such as external DRAM. Furthermore, the vector datapath supports precision gating of its operands to reduce the effective word width and therefore save energy as described in \cite{Moons2016a}. Settings such as the rounding-scheme as well as the fractional-shift of the vector-ALUs can be configured at runtime.

All slots have access to a 32-element wide scalar register file \textit{R} (\mbox{16 bit} per entry). Two large vector register files \textit{VR} and \textit{VRl} of sizes \mbox{16 x 256 bit} and \mbox{12 x 512 bit} respectively are used to provide data to the vector units, thereby acting as an intermediate storage between the on-chip SRAM (DM) and the processor pipeline. The second register file \textit{VRl}, which has double the width of \textit{VR}, is used for vector accumulation. To reduce the multiplexer depth required to access the vector register files, both \textit{VR} and \textit{VRl} are sliced into 4 (\textit{VR0..VR3}) and 3 (\textit{VRl0..VRl2}) sub-regions respectively. While slot 0 can access the complete register files, which is required for data movement and load/stores, the time-critical vector-ALUs only have access to some of the aforementioned sub-regions. Each vector-ALU has an operand fetch and prepare stage that can either broadcast entire vectors to the 4 vector slices within its ALU or generate a permuted version of the input according to a pattern, which is set at runtime.

In addition to the 16 KByte program memory (PM) used to fetch instructions from, ConvAix has access to 128 KByte of dual-ported on-chip SRAM via a custom memory interface. The aforementioned memory is called data memory (DM), which is partitioned into 16 banks of 8 KByte each in order to allow fetches of 2 vectors per cycle (2x256 bit). This is required by the application, since at least one new filter vector and input vector must be loaded each cycle to keep the vector-ALUs busy. To allow seamless transfer of data to/from external memory while the ASIP processes data slices as described in Section \ref{sec_cnn}, there is a simple direct memory access (DMA) engine included in the memory interface. Additionally to the DMA, the line buffer unit has direct access to the memory interface. This allows for simultaneous loads of new IFMap rowchunks while providing (possibly strided) inputs to the vector-ALUs. Using this approach, strided convolutions are executed with minimal cycle overhead.

\section{Results}
\label{sec_results}

\begin{figure}[htb]
\vspace{-0.1cm}
\centering
\includegraphics[scale=0.4]{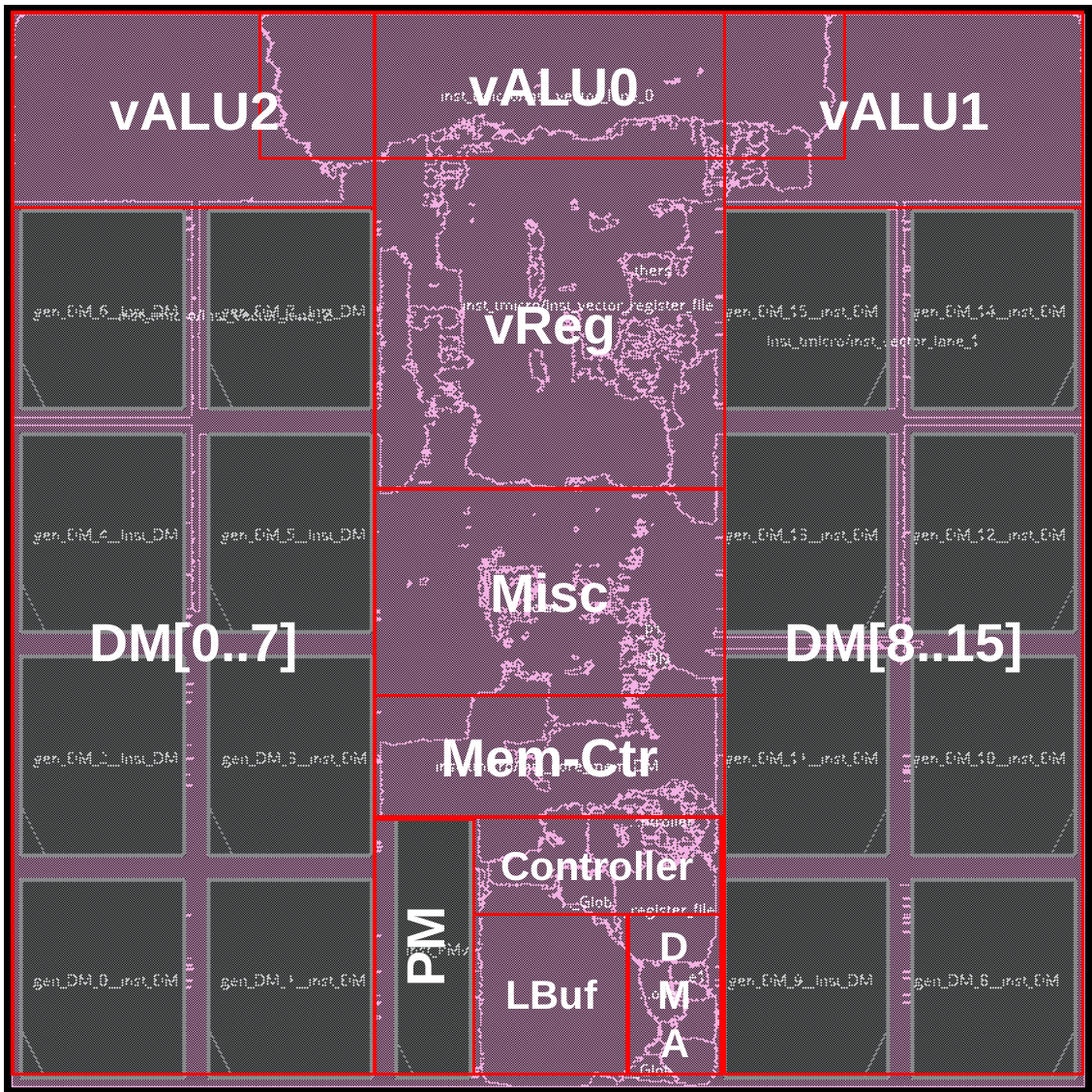}
\caption{Layout view of ConvAix after place \& route.}
\label{fig_layout_view}
\vspace{-0.5cm}
\end{figure}

ConvAix was synthesized and placed \& routed using a TSMC 28nm CMOS technology at 1V nominal supply voltage and standard V\textsubscript{T} for typical conditions (\unit[25]{\degree\ C}). Table \ref{tab_core_spec} summarizes the implementation results, while Fig. \ref{fig_core_area} and Fig. \ref{fig_core_power} present a detailed breakdown of the ASIP's area and power distribution, respectively. The overall layout of the processor is shown in Fig. \ref{fig_layout_view}. All presented power values were obtained by simulating the netlist after place \& route, thereby generating detailed switching activity of the circuit. 

% Table generated by Excel2LaTeX from sheet 'Specification'
\begin{table}[!h]
  \renewcommand{\arraystretch}{0.9}
  \centering
  \caption{\textsc{Processor Specification}}
    \begin{tabular}{|l|l|}
    \hline
    Technology & TSMC 28nm SVT 1P8M \bigstrut\\
    \hline
    Core voltage & 1.0 V \bigstrut\\
    \hline
    Clock frequency & 400 MHz \bigstrut\\
    \hline
    Gate count (logic) & 1293 kGE \bigstrut\\
    \hline
    \multirow{2}[2]{*}{On-Chip SRAM} & 128 KByte (Data) \bigstrut[t]\\
          & 16 KByte (Instruction) \bigstrut[b]\\
    \hline
    \# MAC Units & 192 (3 x 4 x 16) \bigstrut\\
    \hline
    Register Files \& Pipe Registers & 3648 Byte \bigstrut\\
    \hline
    Peak throughput & 153,6 GOP/s \bigstrut\\
    \hline
    \multirow{2}[2]{*}{Arithmetic precision} & \mbox{16 bit} fixed-point \bigstrut[t]\\
          & (+ precision-gating) \bigstrut[b]\\
    \hline
    \end{tabular}%
  \label{tab_core_spec}%
\end{table}%

\begin{table*}[!t]
\vspace{-0.1cm}
    \centering
    \begin{threeparttable}[b]
        \caption{\textsc{Comparison with State-of-the-Art Accelerators}}
        \label{table_comparison}
        \begin{tabular}{r||c|c|c|c|c}
        \multicolumn{1}{r||}{Reference} & Envision \cite{Moons2017a} & \multicolumn{2}{c|}{Eyeriss \cite{Chen2017a}} & \multicolumn{2}{c}{This work (ConvAix)} \bigstrut[b]\\
        \hline
        Technology & 40nm LP (Silicon) & \multicolumn{2}{c|}{65nm LP (Silicon)} & \multicolumn{2}{c}{28nm LP (P\&R)} \bigstrut[t]\\
        Architecture & RISC + MAC Array & \multicolumn{2}{c|}{ASIC} & \multicolumn{2}{c}{ASIP} \\
        Core Voltage [V] & 0.85-0.92 & \multicolumn{2}{c|}{1} & \multicolumn{2}{c}{1} \\
        Gate Count (logic only) [kGE] & 1600  & \multicolumn{2}{c|}{1176} & \multicolumn{2}{c}{1293} \\
        On-Chip SRAM [KByte] & 148   & \multicolumn{2}{c|}{181.5} & \multicolumn{2}{c}{144} \\
        Registers [KByte] & -     & \multicolumn{2}{c|}{11.8} & \multicolumn{2}{c}{3.6} \\
        Clock Frequency [MHz] & 204   & \multicolumn{2}{c|}{200} & \multicolumn{2}{c}{400} \\
        \# MAC Units & 256   & \multicolumn{2}{c|}{168} & \multicolumn{2}{c}{192} \\
        Peak Performance [GOP/s] & 104.5 & \multicolumn{2}{c|}{67.2} & \multicolumn{2}{c}{153.6} \\
        Arithmetic Units & 1-16 bit fixed-pt (scalable) & \multicolumn{2}{c|}{16 bit fixed-pt} & \multicolumn{2}{c}{1-16 bit fixed-pt (scalable)} \bigstrut[b]\\
        \hline
        CNN Model & AlexNet & AlexNet & VGG-16 & AlexNet & VGG-16 \bigstrut\\
        \hline
        Processing Time [ms] & 21.07 & 25.88 & 1251.63 & \textbf{12.60} & \textbf{263.0} \bigstrut[t]\\
        Power Consumption [mW] & 70.1  & 116.8 & 104.8 & 228.8 & 223.9 \\
        Off-Chip I/O [MByte] \tnote{a} & 9.97 \tnote{b}  & \textbf{7.19} \tnote{c} & \textbf{125.8} \tnote{c} & 10.79  \tnote{d} & 208.14 \tnote{d} \\
        MAC Utilization Rate \tnote{e} & 0.61  & \textbf{0.77} & 0.36  & 0.69  & \textbf{0.76} \\
        Area Efficiency [GOP/s/MGE] & 39.73 & 44.01 & 20.85 & \textbf{82.23} & \textbf{90.26} \\
        Energy Efficiency [GOP/s/W] & 815   & 187   & 104   & -     & - \\
        Energy Efficiency @ 28nm/1V [GOP/s/W] \tnote{f} & \textbf{955} & 434   & 242   & 459   & \textbf{497} \bigstrut[b]\\
        \hline
        \end{tabular}%    
        \begin{tablenotes}
       	\item [a] Off-chip I/O for processing batches of size 1
        \item [b] Compressed using Huffman coding
        \item [c] Compressed using run-length coding
        \item [d] Uncompressed
       	\item [e] Ratio of actual and ideal processing time based on 100\% MAC utilization each cycle
        \item [f] Power values were scaled according to the following formula: $P_{scaled} = P_{old} (L_{new} / L_{old}) (V_{DD,new} / V_{DD,old})^2$
        \end{tablenotes}
    \end{threeparttable}
\vspace{-0.35cm}
\end{table*}

Out of the total chip area, the SRAM macro-cells occupy the largest portion at 63\%. As can be seen in Fig \ref{fig_core_area}, the largest area-contributors with regards to the logic-cells are the vector-ALUs, which make up 56\% in total. Regarding the power consumption it can be observed that SRAM data memories together with the register files and the line buffer consume roughly the same amount of power (44.1\%) as the vector-ALUs (44\%). The latter power figures however also include the contribution of pipeline registers and multiplexers within the vector-ALUs.

ConvAix was benchmarked using two widely used CNN models: AlexNet \cite{Krizhevsky2012} and \mbox{VGG-16} \cite{Simonyan2014}. Table \ref{table_comparison} summarizes our results and compares them with two well-known accelerators (Envision \cite{Moons2017a} and Eyeriss \cite{Chen2017a}) targeting the same CNN models. To allow for a fair comparison between all designs, we scaled the energy efficiency values of all architectures to a uniform 28nm technology operating at 1V. The values presented in Table \ref{table_comparison} show overall results across all layers of the respective CNNs with optimized word width for the architectures which provide scalable precision. Furthermore, to increase the fairness of the comparison, the processing times used in Table \ref{table_comparison} do not include the time required for off-chip I/O whenever possible. We hereby aim to eliminate the effect that the I/O bandwidth of the external memories could have on the presented figures.

Due to its comparatively high clock frequency, ConvAix exceeds the other designs in terms of processing speed (1.6x compared to the next fastest in AlexNet and 4.8x for \mbox{VGG-16}) and area efficiency (1.9x for AlexNet and 4.3x for \mbox{VGG-16}). At the same time it maintains a competitive energy efficiency of 459 GOP/s/W on average for AlexNet and 497 GOP/s/W for \mbox{VGG-16}. The average MAC utilization for AlexNet is 8\% lower than that of Eyeriss. This is well expected due to the fact, that the proposed design is software-programmed which always incurs a certain overhead required for control-code. For \mbox{VGG-16} however, ConvAix demonstrates a much higher utilization of 76\% vs. 36\% for Eyeriss. According to the authors of Eyeriss, this is because of added time required for repeatedly ramping up the MAC array. The required off-chip I/O is higher than that of Eyeriss, which can be explained by the lack of a memory compression engine in our design. Calculations using the sparsity-values provided in \cite{Chen2017a} show, that our design would achieve similar total off-chip I/O figures as Eyeriss, if compression was added.

\section{Conclusion}
\label{sec_conclusion}

It was the goal of this work to demonstrate the practical feasibility of a software-programmable architecture with an instruction set that is targeted towards, but not limited to, CNN acceleration. The envisioned architecture, called ConvAix, was implemented in a modern 28nm CMOS technology and evaluated using highly relevant benchmarks. Results show that ConvAix can not only achieve competitive efficiency compared to other less flexible designs, but even surpass them in terms of area efficiency (1.9x for AlexNet, 4.3x for \mbox{VGG-16}) and throughput. Especially for the larger \mbox{VGG-16} model, ConvAix achieves significantly higher utilization (76\% compared to 36\%) and a 4.8x higher processing speed. Incorporating techniques such as dynamic voltage and frequency scaling or memory compression could further improve the efficiency of the presented design. We leave it to future work to investigate this further.

\vspace{-0.1cm}
\section*{Acknowledgment}
This work was supported by the German Federal Ministry
of Education and Research (BMBF) via the PARIS project
(16ES0602) aiming at autonomous driving.

\bibliographystyle{IEEEtran}
\bibliography{IEEEabrv,my_bib}
%\begin{thebibliography}{1}
%\end{thebibliography}

\end{document}